\documentclass[nonacm, sigconf]{acmart}

\setcopyright{none}
\renewcommand\footnotetextcopyrightpermission[1]{}

\usepackage[T1]{fontenc}
\usepackage{lmodern}

\usepackage{balance}

\begin{document}

\title{
Decentralize the feedback infrastructure! 
 }
%\author{}
%\date{}
%\title{It's time for Transparency}
%\author{Pedro Garcia Lopez, Aleksander Slominski, Michael Behrendt, Simon Shillaker, Bernard Metzler\\
%\small {\em  Universitat Rovira i Virgili, IBM, Imperial College London} \\ [2mm]
%\small Submission Type: Research

\author{Pedro Garcia Lopez}
\affiliation{
	\institution{Universitat Rovira i Virgili, Spain}
}
\email{pedro.garcia@urv.cat}

%\author{Aleksander Slominski}
%%\affiliation{
%		\institution{IBM Watson Research, USA}
%}
%\email{aslom@us.ibm.com }

 \date{}

\begin{abstract}

The decentralized architecture of Internet sparkled techno-utopian visions of a virtual freedom space for humanity.  Peer-to-peer systems, collaborative creation (wikipedia), open source software (Linux), universal shared knowledge, and the hopes for disintermediation contributed to this major vision.

However, the reality is bleak: centralization is reigning in the cyberspace, with huge technological corporations controlling our data, and re-intermediation and control are stronger than ever in the so-called "sharing" economy. The Internet is also fragmented by countries, with many states imposing heavy controls to information and communication services.   

The 21st century will witness the major clash between centralization and decentralization in human history. And the major struggle will be around the communication and feedback technologies that will intermediate and govern every interaction in our lives.  

Unlike previous approaches that propose to socialize the feedback infrastructure or to use anti-monopoly laws to break Big Tech companies, in this article we advocate for the  decentralization of the information and communication infrastructure. And the key to this decentralization is the creation of standards enabling interoperability between data platforms. This will in turn produce a true disintermediation from well established technological players and open competition to small third parties. In this article, we sketch such a decentralized open infrastructure including communication, sharing, matchmaking, and reputation services that can be constructed over open source technologies and standards.

\end{abstract}


\maketitle

%-------------------------------------------------------------------------------
\section{Introduction}
%-------------------------------------------------------------------------------

In many aspects of human activity, there has been a continuous struggle between the forces of centralization and decentralization. In history, empires have risen and declined as a consequence of centrifugal and centripetal movements. 

Nowadays, as China emerges, we again live centripetal global movements that lead to a pronouncedly multi-polar decentralized world. But we also witness strong centralization forces in politics, economics and technology. 

China is the example of how centralized authoritarian state capitalism can challenge the more decentralized democratic free market capitalism. And this is shocking, since decentralized markets had already beaten the centralized planning communist economies in the nineties. 

Modern computation and communication technologies are even bringing back the old calculation debate \cite{morozov} to the public sphere.  Can modern Big Data technologies now enable centralized planning not possible in the past century? And even more: Can novel Artificial Intelligence and ubiquitous computing empower centralized states with unlimited social control?

Internet is also becoming more fragmented, as different countries claim sovereignty and establish different rules over their networks like China, Russia, Iran, and many others. Many of these countries have a tight control of their networks, and even the data ecosystem and applications are driven by big local companies like Alibaba, Baidu, VK or Tencent’s WeChat among others.

Whereas the data infrastructure is controlled in China by major state-owned industries, it is mainly managed in the West by major American technology companies like Google, Amazon, Apple or Facebook among others. 

We can also see recent trends towards more state intervention and regulation of the data ecosystem in the West. For example, Morozov recently claimed in “Digital Socialism” to “Socialize the feedback infrastructure” and urge left-wing governments to leverage this important tech infrastructure to improve society. 

Even more liberal options like “Reinventing Capitalism in the Age of Big Data” \cite{reinventing} are claiming for new data markets where the major tech providers share their data with companies and governments.  Another related trend is “Social Media as a public utility” which arguments in favor of government regulation over tech industries, similar to other more established electrical and phone regulated utilities. 

We analyze in this article the implications of centralization and decentralization in human societies and technology. We advocate for a decentralized open data infrastructure relying on human-centered technology systems that respect user’s privacy and freedom while avoiding state’s control for the data infrastructure. 

This does not mean that states must not regulate the data infrastructure. We even advocate that states and federations (like the EU) should collaborate to standardize and even actively support open source data ecosystems and protocols for the digital data infrastructure.

%-------------------------------------------------------------------------------
%\section{Two paths to Transparency}
%\label{sec:figs}
%-------------------------------------------------------------------------------

\section{Centralization Vs Decentralization in History}

The constant struggle between centralization and decentralization can be easily observed during human history. 

Centralization is the basis for the creation of any organized human society from their remote origins. The centralization of control, power, trust and the monopoly of force made possible the creation of huge sophisticated empires. 

However, even the biggest empires required degrees of decentralization and delegation of power, due to the diversity of human settings but also because of communication technologies. The Roman Empire developed an advanced hybrid system with strong centralization of power and standards, but also decentralization to semi-autonomous provinces and kingdoms.  This is also visible in many other heterogeneous and global empires like the Spanish Empire or the British Empire. Furthermore, every empire was always menaced by internal and external centrifugal forces that attempted to disrupt their unity. 

While centralization provided stability and progress in those empires, an excess of centralization could provoke important problems. Excessive centralization may lead to tyranny, lack of freedom, homogeneity and lack of tolerance for diversity, corruption and severe inequalities.   Excessive centralization also hurts innovation, that flourishes with freedom.

As a consequence of those excesses, but also the inherent decline of centralized systems, decentralization provoked revolutions and rebellions, and phases of instability, fragmentation and disaggregation of empires. 

For example, the French revolution finished the absolute centralized power of monarchs and aristocracy, leading to the development of modern states. This also led to a tumultuous period of instability and wars around Europe. 

But probably the biggest clash between centralized and decentralized models has happened in the 20th century, between the socialist centralized state in the URSS and the decentralized market democracies leaded by the US.

It is relevant that the champions of decentralization, 19th century anarchists like Proudhon or Bakunin, already clashed with Marx’s socialists \cite{proudhon2007philosophy,marx1845misery} and broke their relationships soon. The anarchists early alerted about the dangers of centralization and the dictatorship of the proletariat that was defended by Marx’s advocates. 

Western democracies also varied in their decentralized approaches. The USA took a more extreme position, where the state delegated many tasks to the markets and local governments. For example, US education system is very decentralized and responsibility of local entities like towns, which also causes severe inequalities between rich and poor neighborhoods. US also delegated Health services mostly to the market, which in turn provoked again inequality in the access to these expensive services.

European countries followed a more hybrid approach, and social and Christian democrats adopted generous state-supported social systems including education and health, although still following open market-based economies in the shade of the US.

After the demise of the URSS, many could suspect that we reached stability (Fukuyama’s \cite{fukuyama2006end} end of history), but this is no longer true. After the financial crisis in 2008, we are again in the middle of a decentralization wave. Again, decentralization waves mean instability, fragmentation, nationalisms, populisms, protectionism, and global unrest. 

From a Hegelian view of history, this can be considered another step towards the realization of human freedom. We can be tempted to consider that people are unhappy with representative democracies, and thanks to the use of modern social and communication technologies, this can lead to increased freedom, improved radical democracy and decentralization of the power to the people.  

But novel computation and communication technologies are double-edged razors. They can be used to increase social cooperation and communication, but they can also be used to enforce non-democratic centralized control over the population.

In countries like China, centralized technology giants backed by the state have control of all the data ecosystem. However,  in the US the decentralized market has produced giant technology companies that also control most of the data ecosystem. 

The paradox here is that user’s privacy is not guaranteed by either of the systems. In fact, as recently stated by Zuboff in "Surveillance capitalism and the challenge of collective action" \cite{zuboff2019surveillance},  our private data and freedom are  seriously  threatened by surveillance capitalism models in both West and East economies. 

%For example, the state of Utah has granted access to an artificial intelligence company called Banjo to state traffic cameras, CCTV, 911 systems, and state-owned sensitive information. The company will combine all this information with data collected from social media, satellites, and other apps, and use it to feed machine learning algorithms that "detect anomalies" in the real world. As we can see, delegating our privacy to private companies does not solve the problem either.

Let us explore the relationships between decentralization and technology in the last years.

\section{Decentralization and technology }

We could naively consider that technology is following a clear path to decentralization if we analyze the evolution from mainframes to Personal Computers, and to the widespread adoption of mobile phones and devices nowadays. But certainly, this is not true today.

Internet certainly followed a decentralized design based on autonomous systems and internet routing protocols. From being a solely academic network, Internet goes mainstream in the nineties with the advent of the World Wide Web and the universal adoption of Personal Computers (PC). 

Collaborative projects like the Wikipedia, massive oven source projects like Linux, and volunteer computing projects like SETI@HOME fueled the imagination of techno-utopists and libertarians. It is the time when John Perry Barlow presents his “Declaration of Independence of Cyberspace” \cite{barlow1996declaration} claiming sovereignty from existing governments.

In the beginnings of this century Peer-to-Peer (P2P) technologies also flourished with improvements in home network bandwidth and better personal computers. Systems like Napster, Kazaa, and eMule were massively used and only BitTorrent arrived to account to more than 30\% of all Internet traffic.  
 
P2P inspired many efforts to decentralize the Internet like social networks (Diaspora), search engines (Yacy) or storage systems (Wuala). But all of these efforts and P2P applications in general declined or collapsed in front of the competition of Cloud Computing solutions.

Cloud Computing and smart phones represented a big shift towards centralization in the Internet. Nowadays, five cloud service providers control around 60\% of the cloud market share and Amazon  alone  controls  almost  33\%  of it. And consolidation in the Cloud is growing.

Mobile phones heavily rely on Cloud data centers and our personal information is increasingly being stored and processed there with inherent privacy risks. Cloud-based systems from Facebook, Twitter, Google, Dropbox, Netflix or Amazon possess vast amounts of valuable personal data. And recent studies claim that data brokers have thousands of information points from every user in advanced countries.

But, is there hope in other recent technology models like Edge Computing or P2P Blockchains? 

Even the Edge Computing \cite{garcia2015edge} paradigm that advocates to move computations to the edge of the network is mainly following centralized control systems managed from the Cloud.  A clear example is the so called  “Federated learning” \cite {konevcny2016federated}, where a company like Google is centrally orchestrating the execution of machine learning algorithms in user’s mobile devices over their private data (although using differential privacy techniques to provide anonymity).

Blockchains like BitCoin and Ethereum are mainly designed to support their own cryptocurrencies (BTC and Ether) and not as a generic decentralized infrastructure \cite{lopez2019please} for Internet Applications. These networks are slow, and they are consuming enormous amounts of energy in their mining process: Bitcoin consumed in 2019 70TW, which is the approximate annual consumption of a country like Austria.  They are just not designed to decentralize the Internet or to offer data privacy and trust to Internet users.

\section{Property (of personal data) is theft!}

There is an intense debate in the last years around the future of the data economy and the control of the data infrastructure.  In 2017, The Economist published a story titled, "The world's most valuable resource is no longer oil, but data" that has generated a long-lasting impact.

Following this analogy, if data is like oil, then the owner is the first one who drills it. And this is what is exactly happening now in the Internet, where data brokers and tech companies benefit from personal data drilled from users.

Even with recent stringent data protections laws like General Data Protection Regulation (GPDR) or California Consumer Privacy Act (CCPA), many companies still build their business models around personal data and advertisements. 

But how can avoid that our personal data is drilled unpunished in the Internet? The market approach is to privatize data rights: in  “The New Deal on Data” \cite{greenwood2014new}, authors proposed to formalize property rights over data so that users can own the data they produce. This might open new opportunities such as data markets where users may trade their data and obtain benefits out of it. 

The problem of this approach is that, in many cases data has more value when it is aggregated, so a small individual owner will not get much reward, especially for your personal data.  So, selling your soul may not be very profitable either. 

The reality is that there is a wide range of technologies (hardware, software, biological) that can help companies or governments to drill our data without our permission. 

Accessing, trading, or making money using our personal data without our consent is unethical, and basically a theft of our privacy rights.  Governments should protect individuals of abuses to their privacy by third parties.  But either they are weak in front of big multinational companies or authoritarian states themselves are interested in this information. 

\section{Decentralized  Data Infrastructure}

The solutions to the aforementioned challenging problems will not come from authoritarian states or from big technology companies with vested interests.  It will also not arrive through the use of anti-monopoly legislation to split Big Tech companies as suggested by Doctorow \cite{doctorow}. The US cannot allow to fragment their technology champions in front of the massive chinese companies.

It is interesting that the EU is funding research projects like Decentralized Citizens Owned Data Ecosystem \cite{decode}, which aim to create tools that put individuals in control of their personal and shared data. But software projects are live creatures that need continued funding and communities to maintain them.

Tim Berners Lee also tried to create a decentralized web \cite{solid} through personal online data stores (Solid project) ensuring user’s control over data. But the success was very limited, and he now advocates for a social contract for the Web \cite{contract}. As we can see, a project of this size will require the commitment and collaboration of institutions, research institutions and open source companies. 

We advocate here that the solution requires that international entities like the European Union closely collaborate with research institutions and open source companies to create open privacy aware decentralized data infrastructures. Doctorow \cite{doctorow}  also suggests that the solution can come with interoperability and standards that may open the market and enable decentralized platforms.

And the European Union is a global champion in establishing standards that are later adopted around the world, it is the so called Brussels effect  \cite{bradford2020brussels}. Competing now with well established systems like Facebook, Whatsapp, or Amazon is very difficult, as applications like Diaspora already suffered in the past. When all your contacts are in an existing system, there are few incentives to use another one.  What if open standards enable my small application to talk to all users in the major networks ?

%But the solution is more about software than about hardware and data centers. It is not necessary nor realistic to replace American cloud providers with new European Clouds to ensure data sovereignty.   In this line, the European project GAIA-X \cite{gaiax}  has already raised concerns between companies like Amazon or Microsoft.

%As computing resources become standardized utilities in the next years (like the electric grid), it is perfectly possible to build on top of existing clouds privacy-aware decentralized technologies where users retake control of their information. 

What areas are prone to standardization in order to boost a novel decentralized Internet ?
Let’s study some alternatives for this data infrastructure.

\subsection{Personal data repositories}

In the next years, every citizen should have storage and computing resources in the cloud. Our data is now dispersed between government organizations, hospitals, schools, banks and companies. In a few years, we will have a unified digital P.O box in the Cloud where we will be able to control all our data interactions and retake control of our information. Computational resources like storage and computation should be a right for every citizen guaranteed by our taxes.

Governments should provide federated and decentralized identity mechanisms so that our data interchanges are safeguarded. Standardized open protocols should enable user communication and data exchanges among different data repositories and user’s digital PO Boxes.

The required technologies to build this decentralized data infrastructure are already available. The P2P research community advocated for hybrid edge-centric computing technologies \cite{garcia2015edge} that rely on cloud computing resources but decentralizing control to the users. Such human-centered decentralized designs clearly contrast with the traditional master-worker edge computing models like federated learning.

And permissioned open source blockchain technologies can create autonomous systems guaranteeing trust and decentralized entity to users in the Internet. Such autonomous systems should be maintained by local institutions and governments, in line with EU efforts like the EU Blockchain Observatory \& Forum.

\subsection{Shared data repositories}

According to Wikipedia, the commons is the cultural and natural resources accessible to all members of a society, including natural materials such as air, water, and a habitable earth. These resources are held in common, not owned privately.

If we consider unethical to let companies trade with our education or our health for their own benefit, it must be equally unethical to let companies appropriate and benefit from our personal and collective data.

Nobel prize economist Elinor Ostrom in her book “Governing the Commons” \cite{ostrom1990governing} examined the use of collective action, trust, and cooperation in the management of common resources.  We can learn from these models to establish new governance models for the decentralized data infrastructure.

A first step in that direction is the creation of data trusts \cite{o2019data}, where shared data can be delegated to a trusted third party, that can represent and defend the interest of the group. But again, depending on the governance and rules of the data trust we can move towards data markets and sheer profit, or towards ethical governance and social benefits for the group.

The decentralized data infrastructure should provide mechanisms for privacy-aware data sharing, but also other group-based services like reputation and matchmaking.

\subsection{Reputation is the new oil}

As stated by Hendrixs \cite{hendrikx2015reputation}, “Reputation is a tool to facilitate trust between entities, as it increases the efficiency and effectiveness of online services and communities”. Online reputation systems will play an essential role in the future, since they allow trustworthy interactions based on real feedback from other users.  

Gloria Origgi \cite{origgi2019reputation} outlines the importance of reputation and its relationship with trust in the information rich societies we live today. She claims that reputation is needed since it provides second-order epistemology. We are then able to check the indirect reliability of the information like its authority, its reputation.

Reputation systems will become essential services provide trust to users and third parties in the Internet. A number of online platforms already include wide-spread reputation systems like eBay, Amazon, TripAdvisor, Uber, or Google among many others.

But Howard Rheingold states that online reputation systems are "computer-based technologies that make it possible to manipulate in new and powerful ways an old and essential human trait". Reputation scores can become a powerful peer pressure mechanism able to classify or rank individuals or organizations in five stars scores. 

As brilliantly portrayed by dystopic “Nosedive” Black Mirror’s episode, reputation can become a stringent social strait jacket creating new social classes. If the system is centrally controlled, like Chinese Social Credit system, the dangers of misuse are even greater.  

Reputation is another important service that cannot be delegated to the market or companies. Such an important service cannot be in the hands of a private company, nor in the sole hands of the state. We need here again decentralized federations ensuring trustworthy and untampered reputation models thanks to the participation of many heterogeneous participants.

Collaborative filtering and recommender systems are closely related to reputation systems. Collaborative filtering leverage user’s feedback from users to find similarities between them and recommend them services or products.  For example, this can be used to recommend as music, films, or even persons in matchmaking dating services.

The data infrastructure can be used as a matchmaking service for producers and consumers, but also to create novel mechanisms of social cooperation that may transcend markets and money.

Successful technology companies are mainly intermediaries, like Airbnb, Uber, Booking or even Amazon among many others. Silicon Valley has played an innovative role in creating all this technology intermediaries for all kind of services.

If we can devise decentralized and trustworthy matchmaking services, coupled with reliable reputation systems, we could considerably improve open markets and accommodate without friction supply and demand in many fields of human activity.
\newpage

\section{Decentralize the feedback Infrastructure !}

The 21st century will witness the major clash between centralization and decentralization in human history. And the major struggle will be around the communication technologies that will intermediate and govern every interaction in our lives.  

The fight now is not between the centralized state and the decentralized markets. In the east, China has produced state-controlled technology giants that represent a threat for freedom and privacy. But in the west, decentralized markets (USA) have also consolidated all-powerful technology giants that govern our digital lives. Both alternatives present a dim centralized future that threatens our freedom.

The solution is not to socialize the feedback infrastructure, if we understand socializing as giving the state control of that infrastructure. Centralization is dangerous for our freedom, and the left should not repeat the mistakes of the past. 

The left must look back to decentralized proposals: like Proudhon and Ostrom federations \cite{ostrom1990governing}, to self-organization and cooperation, to leverage reputation as a fundamental social incentive. This is not a discussion between centralized state vs decentralized markets, this is about decentralized cooperation facilitated by the state, and about radical democracy.

Yuval Noah Harari in “the world after coronavirus” \cite{covid} states that we face two particularly important choices:  totalitarian surveillance vs citizen empowerment and nationalist isolation vs global solidarity. The coronavirus crisis is pushing strongly towards centralization and strong state intervention inside the states, and non-cooperative decentralization and protectionism between states.  All of this even makes more acute the technology problems portrayed in this article.

The solution is to decentralize the data and communications infrastructure. But this requires the active collaboration and resources of the state. Markets alone do not have the incentives to build this platform.  

The European Union should invest research and technology funds in order that research groups (technology, social sciences) and open source companies can work together to standardize the decentralized data infrastructure. The technologies and knowledge are nowadays available, but the magnitude of the project requires investments and commitment from the states. 

Once in place, the benefits of decentralization would be obvious for states and local communities. Distributed federations would sprout around the world thanks to open protocols, like the Internet exploded in the past as a universal medium.  

The decentralized data infrastructure would still rely on major data center resources, but as controlled utilities like the power grid infrastructure. And computing resources could be used irrespective from the provider, being able to move data or software among different computing resources.

This infrastructure can then be the necessary lever to change the society and increase decentralization of power and democratic participation. As Ursula K Leguin brightly exposed:

\textit{We live in capitalism. Its power seems inescapable. So did the divine right of kings. Any human power can be resisted and changed by human beings. Resistance and change often begin in art, and very often in our art, the art of words.}

{ \balance
{

\bibliographystyle{ACM-Reference-Format}}
\bibliography{ccr}


\begin{thebibliography}{21}

%%% the Web address.

\ifx \showCODEN    \undefined \def \showCODEN     #1{\unskip}     \fi
\ifx \showDOI      \undefined \def \showDOI       #1{#1}\fi
\ifx \showISBNx    \undefined \def \showISBNx     #1{\unskip}     \fi
\ifx \showISBNxiii \undefined \def \showISBNxiii  #1{\unskip}     \fi
\ifx \showISSN     \undefined \def \showISSN      #1{\unskip}     \fi
\ifx \showLCCN     \undefined \def \showLCCN      #1{\unskip}     \fi
\ifx \shownote     \undefined \def \shownote      #1{#1}          \fi
\ifx \showarticletitle \undefined \def \showarticletitle #1{#1}   \fi
\ifx \showURL      \undefined \def \showURL       {\relax}        \fi
% invisible to TeX
\providecommand\bibfield[2]{#2}
\providecommand\bibinfo[2]{#2}
\providecommand\natexlab[1]{#1}
\providecommand\showeprint[2][]{arXiv:#2}

\bibitem[\protect\citeauthoryear{Barlow}{Barlow}{1996}]%
        {barlow1996declaration}
\bibfield{author}{\bibinfo{person}{John~Perry Barlow}.}
  \bibinfo{year}{1996}\natexlab{}.
\newblock \bibinfo{title}{Declaration of Independence for Cyberspace}.
\newblock
\newblock
\urldef\tempurl%
\url{https://www.eff.org/cyberspace-independence}
\showURL{%
\tempurl}


\bibitem[\protect\citeauthoryear{Bradford}{Bradford}{2020}]%
        {bradford2020brussels}
\bibfield{author}{\bibinfo{person}{Anu Bradford}.}
  \bibinfo{year}{2020}\natexlab{}.
\newblock \bibinfo{booktitle}{\emph{The Brussels effect: How the European Union
  rules the world}}.
\newblock \bibinfo{publisher}{Oxford University Press, USA}.
\newblock


\bibitem[\protect\citeauthoryear{Comission}{Comission}{2019}]%
        {decode}
\bibfield{author}{\bibinfo{person}{European Comission}.}
  \bibinfo{year}{2019}\natexlab{}.
\newblock \bibinfo{title}{Decentralized Citizens Owned Data Ecosystem
  (DECODE)}.
\newblock
\newblock
\urldef\tempurl%
\url{https://decodeproject.eu/}
\showURL{%
\tempurl}


\bibitem[\protect\citeauthoryear{Doctorow}{Doctorow}{1996}]%
        {doctorow}
\bibfield{author}{\bibinfo{person}{Cory Doctorow}.}
  \bibinfo{year}{1996}\natexlab{}.
\newblock \bibinfo{title}{How to destroy surveeillance capitalism}.
\newblock
\newblock
\urldef\tempurl%
\url{https://onezero.medium.com/how-to-destroy-surveillance-capitalism-8135e6744d59}
\showURL{%
\tempurl}


\bibitem[\protect\citeauthoryear{Fukuyama}{Fukuyama}{2006}]%
        {fukuyama2006end}
\bibfield{author}{\bibinfo{person}{Francis Fukuyama}.}
  \bibinfo{year}{2006}\natexlab{}.
\newblock \bibinfo{booktitle}{\emph{The end of history and the last man}}.
\newblock \bibinfo{publisher}{Simon and Schuster}.
\newblock


\bibitem[\protect\citeauthoryear{Garcia~Lopez, Montresor, Epema, Datta,
  Higashino, Iamnitchi, Barcellos, Felber, and Riviere}{Garcia~Lopez
  et~al\mbox{.}}{2015}]%
        {garcia2015edge}
\bibfield{author}{\bibinfo{person}{Pedro Garcia~Lopez},
  \bibinfo{person}{Alberto Montresor}, \bibinfo{person}{Dick Epema},
  \bibinfo{person}{Anwitaman Datta}, \bibinfo{person}{Teruo Higashino},
  \bibinfo{person}{Adriana Iamnitchi}, \bibinfo{person}{Marinho Barcellos},
  \bibinfo{person}{Pascal Felber}, {and} \bibinfo{person}{Etienne Riviere}.}
  \bibinfo{year}{2015}\natexlab{}.
\newblock \bibinfo{title}{Edge-centric computing: Vision and challenges}.
\newblock
\newblock


\bibitem[\protect\citeauthoryear{Greenwood, Stopczynski, Sweatt, Hardjono,
  Pentland, Lane, and Nissenbaum}{Greenwood et~al\mbox{.}}{2014}]%
        {greenwood2014new}
\bibfield{author}{\bibinfo{person}{Daniel Greenwood},
  \bibinfo{person}{Arkadiusz Stopczynski}, \bibinfo{person}{Brian Sweatt},
  \bibinfo{person}{Thomas Hardjono}, \bibinfo{person}{Alex Pentland},
  \bibinfo{person}{J Lane}, {and} \bibinfo{person}{H Nissenbaum}.}
  \bibinfo{year}{2014}\natexlab{}.
\newblock \showarticletitle{The new deal on data: A framework for institutional
  controls}.
\newblock \bibinfo{journal}{\emph{Privacy, Big Data, and the public good:
  Frameworks for engagement}}  \bibinfo{volume}{1} (\bibinfo{year}{2014}),
  \bibinfo{pages}{192--210}.
\newblock


\bibitem[\protect\citeauthoryear{Harari}{Harari}{2020}]%
        {covid}
\bibfield{author}{\bibinfo{person}{Yuval~Noah Harari}.}
  \bibinfo{year}{2020}\natexlab{}.
\newblock \bibinfo{title}{The world after coronavirus}.
\newblock
\newblock
\urldef\tempurl%
\url{https://www.ft.com/content/19d90308-6858-11ea-a3c9-1fe6fedcca75}
\showURL{%
\tempurl}


\bibitem[\protect\citeauthoryear{Hendrikx, Bubendorfer, and Chard}{Hendrikx
  et~al\mbox{.}}{2015}]%
        {hendrikx2015reputation}
\bibfield{author}{\bibinfo{person}{Ferry Hendrikx}, \bibinfo{person}{Kris
  Bubendorfer}, {and} \bibinfo{person}{Ryan Chard}.}
  \bibinfo{year}{2015}\natexlab{}.
\newblock \showarticletitle{Reputation systems: A survey and taxonomy}.
\newblock \bibinfo{journal}{\emph{J. Parallel and Distrib. Comput.}}
  \bibinfo{volume}{75} (\bibinfo{year}{2015}), \bibinfo{pages}{184--197}.
\newblock


\bibitem[\protect\citeauthoryear{Kone{\v{c}}n{\`y}, McMahan, Yu, Richt{\'a}rik,
  Suresh, and Bacon}{Kone{\v{c}}n{\`y} et~al\mbox{.}}{2016}]%
        {konevcny2016federated}
\bibfield{author}{\bibinfo{person}{Jakub Kone{\v{c}}n{\`y}},
  \bibinfo{person}{H~Brendan McMahan}, \bibinfo{person}{Felix~X Yu},
  \bibinfo{person}{Peter Richt{\'a}rik}, \bibinfo{person}{Ananda~Theertha
  Suresh}, {and} \bibinfo{person}{Dave Bacon}.}
  \bibinfo{year}{2016}\natexlab{}.
\newblock \showarticletitle{Federated learning: Strategies for improving
  communication efficiency}.
\newblock \bibinfo{journal}{\emph{arXiv preprint arXiv:1610.05492}}
  (\bibinfo{year}{2016}).
\newblock


\bibitem[\protect\citeauthoryear{Lee}{Lee}{2019}]%
        {contract}
\bibfield{author}{\bibinfo{person}{Tim~Berners Lee}.}
  \bibinfo{year}{2019}\natexlab{}.
\newblock \bibinfo{title}{Contract for the Web}.
\newblock
\newblock
\urldef\tempurl%
\url{https://contractfortheweb.org/}
\showURL{%
\tempurl}


\bibitem[\protect\citeauthoryear{Lopez, Montresor, and Datta}{Lopez
  et~al\mbox{.}}{2019}]%
        {lopez2019please}
\bibfield{author}{\bibinfo{person}{Pedro~Garcia Lopez},
  \bibinfo{person}{Alberto Montresor}, {and} \bibinfo{person}{Anwitaman
  Datta}.} \bibinfo{year}{2019}\natexlab{}.
\newblock \showarticletitle{Please, do not decentralize the Internet with
  (permissionless) blockchains!}. In \bibinfo{booktitle}{\emph{2019 IEEE 39th
  International Conference on Distributed Computing Systems (ICDCS)}}. IEEE,
  \bibinfo{pages}{1901--1911}.
\newblock


\bibitem[\protect\citeauthoryear{Mansour, Sambra, Hawke, Zereba, Capadisli,
  Ghanem, Aboulnaga, and Berners-Lee}{Mansour et~al\mbox{.}}{2016}]%
        {solid}
\bibfield{author}{\bibinfo{person}{Essam Mansour}, \bibinfo{person}{Andrei~Vlad
  Sambra}, \bibinfo{person}{Sandro Hawke}, \bibinfo{person}{Maged Zereba},
  \bibinfo{person}{Sarven Capadisli}, \bibinfo{person}{Abdurrahman Ghanem},
  \bibinfo{person}{Ashraf Aboulnaga}, {and} \bibinfo{person}{Tim Berners-Lee}.}
  \bibinfo{year}{2016}\natexlab{}.
\newblock \showarticletitle{A demonstration of the solid platform for social
  web applications}. In \bibinfo{booktitle}{\emph{Proceedings of the 25th
  International Conference Companion on World Wide Web}}.
  \bibinfo{pages}{223--226}.
\newblock


\bibitem[\protect\citeauthoryear{Marx}{Marx}{1845}]%
        {marx1845misery}
\bibfield{author}{\bibinfo{person}{K Marx}.} \bibinfo{year}{1845}\natexlab{}.
\newblock \bibinfo{title}{The Misery of Philosophy: A Response to the
  Philosophy of Misery of Proudhon}.
\newblock
\newblock


\bibitem[\protect\citeauthoryear{Mayer-Sch{\"o}nberger and
  Ramge}{Mayer-Sch{\"o}nberger and Ramge}{2018}]%
        {reinventing}
\bibfield{author}{\bibinfo{person}{Viktor Mayer-Sch{\"o}nberger} {and}
  \bibinfo{person}{Thomas Ramge}.} \bibinfo{year}{2018}\natexlab{}.
\newblock \bibinfo{booktitle}{\emph{Reinventing capitalism in the age of big
  data}}.
\newblock \bibinfo{publisher}{Basic Books}.
\newblock


\bibitem[\protect\citeauthoryear{Morozov}{Morozov}{2019}]%
        {morozov}
\bibfield{author}{\bibinfo{person}{Evgeny Morozov}.}
  \bibinfo{year}{2019}\natexlab{}.
\newblock \showarticletitle{Digital Socialism? The Calculation Debate in the
  Age of Big Data}.
\newblock \bibinfo{journal}{\emph{New Left Review}} \bibinfo{number}{116}
  (\bibinfo{year}{2019}), \bibinfo{pages}{33--67}.
\newblock


\bibitem[\protect\citeauthoryear{O'hara}{O'hara}{2019}]%
        {o2019data}
\bibfield{author}{\bibinfo{person}{Kieron O'hara}.}
  \bibinfo{year}{2019}\natexlab{}.
\newblock \showarticletitle{Data trusts: Ethics, architecture and governance
  for trustworthy data stewardship}.
\newblock  (\bibinfo{year}{2019}).
\newblock


\bibitem[\protect\citeauthoryear{Origgi}{Origgi}{2019}]%
        {origgi2019reputation}
\bibfield{author}{\bibinfo{person}{Gloria Origgi}.}
  \bibinfo{year}{2019}\natexlab{}.
\newblock \bibinfo{booktitle}{\emph{Reputation: What it is and why it
  Matters}}.
\newblock \bibinfo{publisher}{Princeton University Press}.
\newblock


\bibitem[\protect\citeauthoryear{Ostrom}{Ostrom}{1990}]%
        {ostrom1990governing}
\bibfield{author}{\bibinfo{person}{Elinor Ostrom}.}
  \bibinfo{year}{1990}\natexlab{}.
\newblock \bibinfo{booktitle}{\emph{Governing the commons: The evolution of
  institutions for collective action}}.
\newblock \bibinfo{publisher}{Cambridge university press}.
\newblock


\bibitem[\protect\citeauthoryear{Proudhon}{Proudhon}{2007}]%
        {proudhon2007philosophy}
\bibfield{author}{\bibinfo{person}{Pierre-Joseph Proudhon}.}
  \bibinfo{year}{2007}\natexlab{}.
\newblock \bibinfo{booktitle}{\emph{The philosophy of misery}}.
\newblock \bibinfo{publisher}{Cosimo, Inc.}
\newblock


\bibitem[\protect\citeauthoryear{Zuboff}{Zuboff}{2019}]%
        {zuboff2019surveillance}
\bibfield{author}{\bibinfo{person}{Shoshana Zuboff}.}
  \bibinfo{year}{2019}\natexlab{}.
\newblock \showarticletitle{Surveillance capitalism and the challenge of
  collective action}. In \bibinfo{booktitle}{\emph{New labor forum}},
  Vol.~\bibinfo{volume}{28}. SAGE Publications Sage CA: Los Angeles, CA,
  \bibinfo{pages}{10--29}.
\newblock


\end{thebibliography}

\end{document}